**Sodium Allostasis: A New Paradigm for Understanding Cardiovascular Volume Accommodation**

**James Brian Byrd, MD, MS**

Division of Cardiovascular Medicine, Department of Internal Medicine, University of Michigan Medical School

Short title: Sodium Allostasis: A New Paradigm

Corresponding Author:

James Brian Byrd, MD, MS

jbbyrd@med.umich.edu

5570C Medical Science Research Building II

1150 W. Medical Center Drive

Ann Arbor, MI 48109

Total Word Count: 6,934

**Abstract**

Arthur Guyton's classic pressure-natriuresis model posits that dietary sodium challenges induce a transient expansion of blood volume that the kidneys rapidly rectify to restore a strict homeostatic baseline, reducing cardiac output to baseline values despite continued higher sodium diet. While this model remains a cornerstone of medical education, it was established in animal models that included surgical reductions in renal mass. Decades of direct human evidence, including long-term balance and spaceflight simulation studies, show that healthy individuals exhibit substantial, persistent plasma volume expansion and sustained cardiac output elevation during prolonged high-sodium intake with little effect on the mean arterial blood pressure.

To reconcile these empirical inconsistencies, we propose "sodium allostasis" as an alternative regulatory framework. We argue that sodium regulation operates via two distinct mechanisms: a strict concentration homeostasis that defends plasma sodium levels (~140 mEq/L) via dilution, and an allostatic accommodation that can sustain a persistently expanded blood volume and higher total body sodium mass.

This paradigm fundamentally reframes salt sensitivity. Rather than a primary defect in renal sodium excretion, salt sensitivity reflects a failure of the vasculature to accommodate expanded plasma volume. Furthermore, human and animal data reveal that this chronic allostatic state carries a hidden, profound metabolic cost, forcing energy-intensive sodium reabsorption into poorly oxygenated renal medullary segments and inducing tissue hypoxia independent of blood pressure. Shifting focus from rigid homeostatic volume regulation to vascular accommodation and sodium allostasis provides an accurate physiological foundation for cardiovascular medicine and opens novel, vascular-targeted therapeutic pathways for hypertension and volume disorders.

## Introduction

Arthur C. Guyton's description of the relationship between pressure-natriuresis and blood pressure has profoundly influenced what has been taught to physicians and biomedical researchers for decades. His model proposes that blood pressure is regulated through pressure-dependent changes in renal sodium excretion, creating a negative feedback system with theoretically "infinite gain." He posits that even the smallest perturbation upward in blood pressure leads to additional sodium and volume excretion to such an extent that the blood pressure returns to its starting point.

Guyton's conceptualization was as follows:

> "If extracellular fluid volume and blood volume should become slightly elevated, the quantitative cascade would cause a tremendous increase in arterial pressure over a number of days. As this pressure rose, it would cause the kidneys to excrete the excess fluid. Thus, the extracellular fluid volume and blood volume both would return to their original control values at the same time that the pressure returned to normal."(1)

He reiterated this reasoning in other writings, as well,(2) and this model continues to be widely taught.

These statements are consistent with Guyton's findings in a study of dogs approximately 70% of whose kidney mass had been removed.(3) Using the Evans blue method of plasma volume measurement, Guyton found that in these partially nephrectomized dogs, a 17.5% increase in blood volume was rectified by day 20 during ongoing dietary sodium administration.(3) The claim that volume normalizes and does not persistently expand in healthy individuals remains current, appearing both in the standard medical physiology teaching text and in the primary research literature. The current (fifteenth) edition of Guyton and Hall's Textbook of Medical Physiology states that, absent impaired kidney function or excess antinatriuretic hormones, increased salt intake "usually does not increase arterial pressure much because the kidneys rapidly eliminate the excess salt, and blood volume is hardly altered."(4) Direct measurement in healthy humans contradicts the claim that blood volume is hardly altered.

Results of human salt-loading studies are inconsistent with a pivotal prediction of this model: restoration of blood volume to within 5% of its prior quantity despite ongoing high-sodium diet.(2) Beyond salt-loading studies, the physiology of pregnancy is an example of a persistent 40-50% increase in plasma volume(5) that neither gets quickly rectified, nor causes hypertension in most pregnancies. That's true despite increased circulating sodium mass and increased glomerular filtration rate. Thus, a sustained increase in filtered sodium load and reabsorption can be accommodated by the vascular system in a way that yields a net *decreased* blood pressure. In the pathological context, primary aldosteronism demonstrates that across a wide range of blood pressures including normotension,(6) long-term plasma volume expansion is not only possible in humans but can cause an increase in glomerular filtration rate ("hyperfiltration"), translating into increased filtered sodium load and energy expenditure on sodium reabsorption. Resolution of primary aldosteronism by surgical adrenalectomy is followed by a decrease in glomerular filtration rate demonstrating this fact clearly. In contrast, primary hypertension is a state of relatively low plasma volume.(7) Taken together, these findings suggest the need for an alternative framework that better explains observed normal physiological responses to dietary sodium. This new framework will force a productive reimagining of how hypertension, primary aldosteronism, heart failure, and other pathological conditions operate, with greater attention to the interdependence of the heart, vasculature, and kidney. While sodium homeostasis well deserves the attention it has received in 15,700 works indexed in Google Scholar, there is a need for a new term beyond “sodium homeostasis” to distinguish those aspects of sodium metabolism that *are* homeostatic from aspects that involve adaptations to new and persistently changed levels.

**Normal Physiology versus Adaptations to Partial Nephrectomy**

A critical examination of Guyton's foundational work reveals a profound methodological contradiction that has shaped cardiovascular education for decades. Guyton's experimental models required surgical destruction of kidney tissue specifically because healthy animals were remarkably resilient to salt-induced hypertension.(8)

As documented by Guyton, normal animals rarely develop hypertension with dietary sodium challenges alone.(8) Guyton surgically removed about 70% of the kidney mass to create salt-sensitive models. Whereas Guyton pursued the salt-sensitive models, others explored the resilience of the unoperated animals: normal physiology, in other words. Guyton’s finding of the resolution of plasma volume expansion at 20 days into the salt-loading of dogs after partial nephrectomy(3) stands in stark contrast to another group’s finding in normal, unoperated dogs. Gupta et al. showed that volume expansion persists for 6-10 weeks during high-sodium diet in normal dogs.(9)

In normotensive humans, even across a 30-fold range of sodium intake (10 to 300 mEq/day), Luft et al. observed an 18-fold increase in sodium excretion with no significant change in systolic or diastolic blood pressure, while cardiac index increased and calculated peripheral vascular resistance decreased.(10) In our own dietary sodium crossover study with a comparable sodium range, mean arterial pressure and diastolic blood pressure were lower during the high-sodium condition despite a 1.4 kg weight gain, confirming expansion of extracellular fluid volume.(11) Taken in conjunction with studies showing that sodium-related expansion of extracellular fluid volume is accompanied by expanded plasma volume, our finding of expanded volume with decreased blood pressure is consistent with a decreased systemic vascular resistance.

The divergence between this normal physiology and the damaged-kidney models is stark: a fully intact organism gracefully accommodates sodium challenges through volume expansion without hypertension, while damaging the kidneys is sufficient to cause hypertension in this setting. It is worth clarifying that the *sufficiency* of damaging the kidneys to cause hypertension during high sodium diet is not equivalent to the *necessity* of damaging the kidneys to produce hypertension, a possible misunderstanding of Guyton's findings. Evidence of determinants of long-term blood pressure control separable from renal function has been reviewed elsewhere.(12)

**Inconsistencies Between Predicted and Observed Responses**

It has been known since the 1960s that persistently increasing the sodium content of the diet persistently increases total body sodium.(13) This is not surprising. It would be astonishing if a substance could be ingested daily in a doubled amount until a new steady state of input-output balance, but without accumulation of the substance in the body pending the achievement of that input-output balance (i.e., pending steady state). For example, it is widely understood that doubling of the dose of a daily drug results in more drug in the body, even though a new input-output balance will be achieved in due time.

However, the situation with sodium is prone to cause more confusion than for other substances. Sodium has two easily demonstrated homeostatic aspects. First, just like with a drug, after a period of adaptation following a dose change, a new steady state equilibrium occurs in which the amount eliminated from the body is approximately the same as the amount introduced into the body. For clarity, we call that sodium's input-output balance. However, the second homeostatic aspect of sodium's metabolism stands in stark contrast to what happens with a drug, creating great potential for confusion. The plasma concentration of sodium tends to be about the same across a wide range of steady-state dietary sodium intake. For example, the plasma sodium was 140 mEq/L in a group of people habitually eating about 100 mg per day and 142 mEq/L in a group of people habitually eating about 3,500 mg per day.(14) This second aspect of homeostasis masks the accumulation of sodium in an expanded plasma volume that occurs during high-sodium diets.

Guyton wrote that increased fluid intake produces "only minute, usually unmeasurable changes in body fluid volume."(15) Human long-term salt-loading studies directly contradict this prediction, showing substantial weight gains that persist for months. The sodium itself weighs far less than the observed weight gain. The weight gain, however, is most readily explained by the accumulation of water, whether due to water-drinking or mechanisms that increase the retention of ingested water in the body during steady-state increase in dietary sodium intake.

It stands to reason that pressure natriuresis is not inducing a negative input-output sodium balance during the chronic, steady-state phase of a high-sodium diet—after the period of accommodation, which is to say accumulation of sodium and water. If pressure natriuresis were inducing an ongoing sodium and water loss, rather than input-output balance, we would see progressive sodium loss and death. Thus, the question of whether pressure natriuresis regulates volume is relevant to a time-limited window since the effects must have ceased by the time steady-state input-output sodium balance occurs. However, we know that plasma volume acutely expands during sodium loading. Thus, neither in the acute nor in the chronic setting is plasma volume regulated in the sense of tightly controlled by pressure natriuresis, as the physiologists mean by "regulated."

**Human Studies Demonstrate Persistent Volume Expansion and Sustained Increase in Cardiac Output**

That sustained high sodium intake raises total body sodium and holds it elevated, rather than returning it to baseline, was argued by Bonventre and Leaf in 1982 against a model regulating total body sodium around a set point.(16) They held that increased intake produces weight gain and increased total body sodium that persist as long as the higher intake is maintained, with return to lower values contingent on reduced intake rather than on any regulatory restoration. They credit Ludwig in the 19th century with describing sodium retention over roughly three days, followed by re-equalization of excretion with intake at a higher retained level. The persistence of elevated total body sodium is therefore an old observation. The present framework departs from that lineage in several respects developed across this paper: it argues that pressure-natriuresis cannot account for salt resistance in healthy people, developed below; it concerns plasma volume specifically, which expands to a new elevated level with minimal change in blood pressure; it supplies a vocabulary, sodium allostasis, facilitating clear reasoning about homeostatic aspects of sodium (concentration) versus aspects that can arrive and remain at a "new normal" (total circulating sodium mass); and it locates the cost of accommodation and the failure that produces salt sensitivity, treated below. The studies in this section establish the persistence of the plasma-volume expansion; the homeostatic-allostatic distinction is developed in the section that follows.

**Sustained Cardiac Output Elevation at One Year**

The persistence of expanded plasma volume has a direct hemodynamic consequence that has been measured in humans over a full year. Sullivan and Ratts measured cardiac index by echocardiography across sodium depletion and three points during a maintained high-sodium diet followed for 12 months.(17) On a prescribed 10 mEq/day diet cardiac index was 2.32 L/min/m²; after four days at a prescribed 200 mEq/day it was 2.53. It then rose further on the maintained 200 mEq/day intake: 3.11 at 6 months and 3.08 at 12 months ($p<0.025$). The elevation was not confined to the first days of loading; cardiac index was higher at 6 and 12 months than during short-term repletion and remained elevated through the full year. Across the same time span, mean arterial pressure was unchanged (82–84 mm Hg) and total peripheral resistance fell from 1778 to 1282 $dyne \cdot sec \cdot cm^{-5}$. The rise in cardiac index was driven by increased stroke volume at an unchanged heart rate, the higher stroke volume itself resulting from increased left-ventricular end-diastolic volume. The chronic sodium intake was self-administered and verified by 24-hour urinary sodium (144 mEq at 6 months, 171 at 12 months), not controlled feeding; the 6- and 12-month measurements were obtained in seven of the ten enrolled subjects, adequate echocardiograms being unavailable at the later points in three. The return of cardiac output toward baseline observed in the salt-loaded dog with reduced renal mass is not observed in the intact human.

*The Mars Simulation Studies*

Rakova et al. conducted unprecedented long-term balance studies in 10 healthy men during controlled simulations lasting 105-205 days.(18) Subjects were maintained under completely controlled conditions with salt intake varying between 6, 9, and 12 g/day while all other nutrients remained constant. The findings directly contradicted Guyton's predictions. A positive dose-dependent relationship was observed between dietary sodium intake and body weight. Compared to the 6g sodium condition, body weight increased by approximately 2-3 kg in the 12g sodium condition. No weight normalization phase was observed. During free access to water, the body accumulated weight dose-dependently with sodium, and this weight was never normalized despite months of observation

This sustained weight gain represents in part persistent substantial fluid volume expansion - exactly the opposite of Guyton's predicted volume normalization through pressure-natriuresis.

*Additional Relevant Evidence from Human Studies*

An additional 135-day salt-loading study in three normal humans demonstrated sodium accumulation of 2,973–7,324 mEq and weight gains of 5.1–9.3 kg, with no subsequent return to baseline in either measure.(19) Throughout, systolic blood pressure remained essentially unchanged in each subject across the four 33-day study periods (120→119, 110→113, and 113→114 mmHg, respectively), and diastolic pressures were similarly stable. Body weight and total body sodium remained elevated at their new levels for the duration of the study, with no normalization phase, while plasma sodium concentrations remained within the normal range.

Multiple controlled studies using validated plasma volume measurement techniques confirm persistent expansion patterns. Heer et al. conducted two studies, both of which showed that increasing sodium chloride intake increases plasma volume. In a stepwise open-label study raising intake from 220 to 660 mEq/day, plasma volume expanded by 11.4 ± 2.0% ($P < 0.0001$), persisting throughout the high-salt period, with no change in blood pressure ($P > 0.05$). In a separate randomized controlled study, dose-dependent plasma volume increases of up to 13.8 ± 2.9% were confirmed.(20) In addition, there was a positive net sodium input-output balance: the accumulated sodium was not excreted. Notably, total extracellular volume measured by inulin dilution did not change, indicating that the expansion was specifically intravascular, a redistribution of fluid into the plasma compartment rather than generalized extracellular fluid accumulation.(20) El-Sayed and Hainsworth conducted an 8-week double-blind study in people with unexplained syncope, finding that 70% of salt-loaded subjects maintained elevated plasma volumes throughout the entire treatment period, without a significant increase in supine blood pressure despite 8 weeks of volume expansion. The participants who did not expand their plasma volume had more than twice the baseline dietary sodium intake compared to the responders; it is not clear whether they consumed a smaller amount of sodium, or handled it in a different way.(21)

The pattern across studies is remarkably consistent: rapid plasma volume expansion during salt loading, followed by sustained elevation of plasma volume rather than the normalization that Guyton predicted.

### Evidence That Persistent Increase in Plasma Volume and Circulating Sodium Quantity Matters: Metabolic Costs

One could reasonably ask whether the plasma expansion, with nearly identical sodium concentration, caused by high sodium diets causes a substantial increase in workload for the kidney. Although it seems to have been rarely if ever calculated, the increase in circulating plasma sodium mass (in mEq) is easily calculable as the difference between the circulating plasma sodium mass at the higher and lower intakes — that is, $(V_2 \times C_2) - (V_1 \times C_1)$, where V is plasma volume (in liters) and C is plasma sodium concentration (in mEq/L) in each condition. Because the sodium concentration is defended within a narrow range (a homeostatic aspect of sodium), this reduces approximately to the change in plasma volume multiplied by the prevailing concentration. A worked example illustrates the magnitude. Heer et al. reported a baseline plasma volume of 3,027 ± 76 ml at a sodium intake of 220 mEq/day, which increased by 11.4% to approximately 3,372 ml when intake was raised to 660 mEq/day, with no change in plasma sodium concentration.(20) At a plasma sodium concentration of approximately 140 mEq/L, the circulating plasma sodium mass rose from approximately 424 mEq to 472 mEq, an increase of roughly 48 mEq, or about 11%, while the concentration changed by essentially 0%. A larger circulating sodium mass is therefore carried at an essentially unchanged concentration, which is the signature of allostatic rather than homeostatic regulation of total circulating sodium. This 11% increase in circulating sodium mass is a lower bound on the increase in filtered sodium load, because GFR also tends to rise during high-sodium intake.(10, 22) The repeated circulation of these sodium molecules through the kidney throughout the day and the kidney's avid, obligate reabsorption of a high fraction (certainly >95%) of this sodium are well known, as is the cost in ATP of reabsorption of a given amount of sodium. Recent studies demonstrate the metabolic cost of this excess sodium reabsorption.

**Animal Evidence of Metabolic Reprogramming during High-Sodium Diet**

Shimada and colleagues(22) showed the stresses imposed on the kidney by a high-sodium diet. In salt-resistant rats, glomerular filtration rate and renal blood flow increased during high-sodium diet, and tubular oxygen consumption rose progressively over 21 days while the ratio of sodium reabsorbed per unit of oxygen consumed remained constant, indicating that the increased oxygen cost was driven by a larger filtered sodium load rather than by less efficient reabsorption. Shimada et al. also documented metabolic reprogramming during chronic salt loading, including downregulation of efficient energy pathways and sustained inflammatory responses, despite minimal change in blood pressure.(22) Separately, Udwan et al. showed in mice that high dietary sodium redistributes tubular reabsorption from proximal to distal nephron segments, shifting work to less efficient, poorly oxygenated regions.(23) Whether this redistribution is accompanied by an undetected increase in GFR or represents a purely structural reorganization of transport, the net result in both models is increased renal metabolic cost during high-sodium diet. As discussed below, at least in hypertensive humans, glomerular filtration rate and tubular energy expenditure are also increased by high-sodium diet.

In summary, rather than pressure-natriuresis leading to sodium homeostasis as Guyton predicted, kidneys work substantially harder long-term to maintain sodium input-output balance, clear evidence of accommodation rather than regulation back to baseline conditions.

**Direct Human Evidence: BOLD-MRI of the Kidney**

Pruijm et al.(24) used Blood Oxygen Level Dependent Magnetic Resonance Imaging to measure kidney tissue oxygenation in normotensive volunteers during high-sodium versus low-sodium periods. During high-sodium intake, renal medullary oxygenation decreased significantly (medullary R2*: 31.3±0.6 during high sodium versus 28.1±0.8 during low sodium $s^{-1}$, $p<0.05$), indicating tissue hypoxia in oxygen-sensitive regions responsible for sodium concentration. Similar results were obtained in hypertensive volunteers also included in the study. This occurred even though the participants maintained normal plasma sodium concentrations and would have been in sodium input-output balance by day 8, when the BOLD-MRI was performed.

Although the present framework focuses on normal physiology, parallel findings in hypertensive populations are instructive. Limiting their analysis to hypertensive patients, Rossitto et al. recently showed that those consuming a high-sodium diet increase their glomerular filtration rate, which at a constant sodium concentration means that they increase their filtered load of sodium.(25) These authors also saw the increase in tubular energy expenditure implied by the attendant reabsorption of this larger filtered load.

This metabolic stress reveals allostatic, not homeostatic sodium physiology, a distinction described in more detail below. Under low-sodium conditions, most sodium reabsorption occurs efficiently in well-oxygenated proximal tubules. However, processing the massive sodium throughput required during high-salt accommodation forces more energy-intensive reabsorption into poorly perfused medullary segments. The kidney successfully maintains sodium balance and concentration homeostasis, but at considerable metabolic cost, requiring sustained tissue hypoxia to handle the increased workload.

These human data parallel the findings of Shimada and colleagues in rats,(22) demonstrating that sodium accommodation in both salt-resistant rats and healthy humans requires ongoing physiological work rather than the corrective pressure-driven mechanisms Guyton predicted. The kidney achieves sodium balance not through purely homeostatic regulatory mechanisms, but through metabolically expensive adaptations that maintain plasma sodium concentration while accommodating higher plasma volumes, higher circulating sodium quantity, and higher total body sodium levels.

In view of the above considerations, it is not surprising that Guyton's models as implemented computationally underestimated the sustained changes caused by salt loading.(26) Geoffrey Rose memorably pointed out that scientists would not suspect smoking as the cause of lung cancer if everyone in their society smoked 10 cigarettes a day.(27) Likewise, the relatively narrow range of usual sodium consumption in industrialized societies and the paucity of relevant studies in low-sodium populations make it difficult to calculate the full metabolic cost of our high-sodium diets. The current data suggest that ongoing plasma expansion and increased blood sodium content resulting from "ordinary" diets in industrialized societies are sufficient to cause metabolic consequences affecting human health—independent of sodium's effect on blood pressure.

**Sodium Allostasis: A New Framework**

The evidence demands a fundamental reconceptualization of how the cardiovascular system manages the high sodium diets that are ubiquitous in industrialized societies. Beyond some widely recognized homeostatic aspects of sodium metabolism, the system also operates through what we propose to call "sodium allostasis," a regulatory approach that maintains stability through change rather than through restoration to original setpoints.

This is not a distinction without a difference.

**Clearly Distinguishing Homeostasis from Allostasis in Sodium Regulation**

Classical homeostasis implies return to baseline conditions after perturbation. Guyton's model represents the purest expression of homeostatic thinking: any deviation from baseline blood pressure triggers mechanisms that restore both pressure and plasma volume to their original states. However, human data show otherwise.

Two aspects of sodium handling do demonstrate clear homeostatic control. First, plasma sodium concentration remains remarkably stable around 140 mEq/L across wide ranges of dietary sodium intake through dilution mechanisms that adjust total body water content. Second, after an adaptation period, sodium output does closely match sodium input, preventing endless accumulation that would be physiologically catastrophic.

However, other aspects – total amount of circulating sodium and the associated plasma volume that dilutes it, total body sodium, renal sodium reabsorption (mEq per unit of time), oxygen consumed to do that reabsorption - follow allostatic rather than homeostatic principles. During sustained high-sodium intake, plasma volume increases to a new elevated steady state and remains there, not returning to baseline. This represents accommodation to changed circumstances rather than restoration of original conditions. The cardiovascular system thus operates through sodium allostasis, maintaining plasma sodium concentration stability while accepting persistent volume expansion, in addition to very different homeostatic mechanisms also in operation.

**Clinical Implications: Reframing Salt Sensitivity**

This framework fundamentally reframes our understanding of salt sensitivity. Rather than representing enhanced sodium retention, salt sensitivity appears to reflect failed vascular accommodation to the volume expansion that is well demonstrated to accompany high-sodium intake. Salt-resistant or inversely salt sensitive individuals demonstrate the expected allostatic response: they accommodate volume expansion through vasodilation, maintaining normal blood pressure despite the persistently expanded plasma volume circulating a larger number of sodium molecules. Salt-sensitive individuals do not have the same accommodation capacity, experiencing hypertension when volume expands. The human hemodynamic data localize this defect to the vasculature rather than to renal sodium handling, and they do so by what fails to distinguish the two groups as much as by what does. In 157 subjects studied across 10 and 200 mEq/day sodium intakes, Sullivan and Ratts found that salt-sensitive participants, whether normotensive or borderline hypertensive, were characterized by higher forearm vascular resistance during sodium loading and lower venous capacitance, together with relatively suppressed plasma renin activity and aldosterone, while plasma volume did not distinguish salt-sensitive from salt-resistant people(28) That plasma volume is non-discriminating is precisely what the present framework predicts: extracellular and plasma volume expand when sodium intake increases substantially, so volume should not separate those who become hypertensive from those who do not. What separates them is whether the vasculature relaxes, that is, whether total peripheral resistance falls enough to accommodate the elevated output. Salt sensitivity, on these data, is a failure of the negative regulation of vasoconstriction rather than a difference in how much sodium or volume is retained. Kurtz and colleagues independently reached a compatible conclusion from a systematic review of balance data: salt-resistant people do not excrete sodium more rapidly, retain less sodium, or undergo lesser blood volume expansion than salt-sensitive people.(29) Their alternative hypothesis emphasized renal vasodysfunction, but the shared empirical foundation, that differential sodium handling does not explain salt resistance in healthy people, is the same. This localization is mechanistically congruent with the function of SVEP1, a sodium-responsive molecular regulator of vasoconstriction discussed below. Moreover, Evans and Bie have proposed that it is time to wrestle with inconsistencies between the Guytonian paradigm and what is observed in clinical study.(30) We concur, and the present framework offers a specific alternative.

The reconceptualization we propose suggests that approaches that enhance endothelial function, preserve vascular compliance, or support the molecular pathways enabling vasodilation during volume challenges might prove more effective than traditional sodium restriction alone.

Recognition of sodium allostasis opens several critical research directions. First, identifying the molecular determinants of vascular accommodation capacity could reveal why some individuals successfully accommodate volume expansion while others develop hypertension. Second, developing biomarkers for clinical assessment of vascular accommodation would enable personalized approaches to sodium management based on individual physiological capabilities rather than population-wide recommendations. Third, more intensive investigation of vascular accommodation therapeutically through interventions that enhance endothelial function or preserve arterial compliance may prove complementary to existing prevention and treatment approaches in the field of hypertension.

Understanding that normal physiology routinely maintains approximately 10% plasma volume expansion during high-sodium intake, requiring metabolically costly accommodation - fundamentally changes how we should approach salt-sensitive hypertension and volume disorders. Rather than viewing volume expansion as pathological, this framework recognizes it as the normal physiological response to sodium challenges, with pathology occurring only when accommodation mechanisms fail or the costs of the accommodations cause problems of their own.

## Vascular Adaptation to Volume Expansion as a Research Frontier

We recently discovered that plasma SVEP1 was higher during high-sodium diet compared with during low-sodium diet in people who weighed 1.4 kg more during 8 days of high-sodium diet compared with 8 days of low-sodium diet.(11) Meanwhile, their mean arterial blood pressures and diastolic blood pressures were lower on average during the high-sodium diet, and the systolic blood pressure tended to be lower ($P=0.060$). Changes in SVEP1 between the high- and low-sodium diet correlated inversely with changes in blood pressure. SVEP1 acts on integrins $\alpha9\beta1$ and $\alpha4\beta1$ to regulate GPCR-mediated vasoconstriction.(31) A high correlation between changes in NT-ProBNP and SVEP1 ($R=0.8$, $P<0.001$) suggested that volume expansion is involved in the increase in plasma SVEP1 levels. SVEP1 is the most upregulated protein identified thus far in the plasma of women during a normal pregnancy,(32) providing convergent evidence that it is a plasma volume-sensitive molecule that correlates inversely with blood pressure during volume expansion. Others have also observed the inverse salt sensitivity phenotype we found in our dietary sodium crossover clinical trial,(33) although it has not always been discussed even when observed.

Several groups have laid groundwork for lines of vascular adaptation research suggested by the above findings. Peter Bie has demonstrated that changes in natriuresis and changes in blood pressure are not always tightly coupled.(34) There is a separable influence of plasma volume. This makes sense because $MAP = (CO \times SVR) + CVP$, implying that it is easy to find changes in various components of this equation without changes in the other components—due to compensations affecting other components of the equation. A change in plasma volume need not translate into a change in blood pressure triggering pressure natriuresis if vasodilation causes accommodation of the volume without an increase in blood pressure. The arterial tree can decrease the systemic vascular resistance via relaxation of resistance arteries even as the venous compartment of the blood vessels grows in volume considerably due to its high capacitance and proportionally higher fluid content compared to the arteries. Several(35-39) but not all(40) studies of the topic suggest that lymphatic vessels play a role in the relationship between extracellular fluid volume and arterial pressure during high-sodium diet.

Work by Oberleithner and colleagues has helped delineate the role of the endothelial glycocalyx as a buffer for increased sodium intake, standing between the dietary sodium load and the endothelial cells.(41) There have been increasing calls for re-evaluation of how we envision the issues described above,(42) and in proposing this framework of sodium allostasis, this analysis provides an improved coherence and more integrated explication of the higher-order issues at hand. A key modern explanation of salt sensitivity is that the expanded ECV during high-sodium diet requires a higher blood pressure in some people to effect the pressure natriuresis needed for sodium balance. This is conceptually similar to saying that some people's vasculature accommodates a larger volume without an increase in blood pressure. However, the latter statement hews more closely to the observation showing humans can and do accommodate additional plasma volume long-term by reducing systemic vascular resistance, sometimes with a net *decrease* in blood pressure.

**Conclusion**

Guyton's pressure-natriuresis model faces decisive empirical refutation. The model predicted rapid volume normalization through pressure-mediated mechanisms, but human studies document persistent volume expansion lasting months. Most importantly, the foundational experiments required surgical kidney damage precisely because normal physiology was too resilient - yet these pathophysiological models became the basis for teaching normal cardiovascular physiology.

Human physiology appears to operate through sodium allostasis—with blood sodium concentration tightly regulated while blood volume is accommodated through metabolically expensive but effective mechanisms. This framework better explains observed phenomena and provides improved foundations for cardiovascular therapeutics.

**Funding**

JBB is funded by NIH/NHLBI R01HL164678.

**Glossary**

**Accommodation** - The physiological process of adapting to and maintaining function at a new, elevated level (e.g., higher sodium content) rather than returning to baseline conditions.

**Allostasis** - The process of achieving stability through physiological change and adaptation to new conditions, allowing for different set points depending on circumstances. Distinguished from homeostasis, which maintains fixed set points.

**BOLD-MRI** - Blood Oxygen Level Dependent Magnetic Resonance Imaging technique that measures tissue oxygenation by detecting changes in blood oxygen levels.

**Concentration Homeostasis** - The maintenance of stable plasma sodium concentration (~140 mEq/L) regardless of daily dietary sodium intake, total body sodium content, or plasma volume changes.

**Infinite Gain** - Guyton's theoretical concept describing a feedback system so powerful that any blood pressure increase triggers proportionally massive sodium and water excretion, theoretically returning pressure to its starting point.

**Inverse salt sensitivity** - A situation in which blood pressure decreases significantly in response to high sodium intake. The repeatability of the trait over time is poorly studied, but that it exists demonstrates that the range of possible reactions to a blood volume increase is wide.

**Sodium Input-Output Balance** - A steady state where the amount of sodium consumed equals the amount eliminated, irrespective of total body sodium content. Contrast with the common, vague term "sodium balance," which might be misunderstood to mean homeostasis of total body sodium.

**Pressure-Natriuresis** - Pressure-dependent changes in renal sodium excretion, creating a negative feedback loop, and proposed by Guyton to explain salt resistance.

**Salt Sensitivity** - A situation (very possibly a long-term condition) in which the blood pressure increases significantly in response to high sodium intake, contrasted with salt resistance in which blood pressure remains stable or inverse salt sensitivity in which blood pressure decreases.

**Sodium Allostasis** - The proposed new paradigm where plasma sodium concentration remains stable (~140 mEq/L) through dilution mechanisms while total body sodium and plasma volume persistently increase to accommodate high dietary sodium intake.

**Total Body Sodium** - The complete amount of sodium in the body, which can increase substantially even while plasma sodium concentration remains unperturbed through dilution.

**Vascular Accommodation** - The ability of the cardiovascular system to adapt to increased plasma volume through enhanced compliance, vasodilation, and lymphangiogenesis without proportional pressure increases.